\documentclass[aps,prl,preprint,superscriptaddress,bibliography]{revtex4-1}
\usepackage{amsmath,amssymb,graphicx,bm}
\usepackage{xcolor}
\usepackage[colorlinks=true,urlcolor=blue,linkcolor=blue,citecolor=blue,bookmarks=false]{hyperref}
\usepackage{ulem}
\usepackage{placeins}
\usepackage{physics}
\usepackage[utf8]{inputenc} 

\newcommand{\beginsupplement}{%
        \setcounter{table}{0}
        \renewcommand{\thetable}{S\arabic{table}}%
        \setcounter{figure}{0}
        \renewcommand{\thefigure}{S\arabic{figure}}%
     }

\usepackage{times}

\usepackage{float}

\begin{document} 
\title{Tunable gigahertz dynamics of low-temperature skyrmion lattice in a chiral magnet} 

\keywords{}

\author{Oscar Lee}\email{s.lee.14@ucl.ac.uk}
\affiliation{London Centre for Nanotechnology, University College London, United Kingdom}

\author{Jan Sahliger}
\affiliation{Physik-Department, Technische Universit\"at M\"unchen, D-85748 Garching, Germany}

\author{Aisha Aqeel}
\affiliation{Physik-Department, Technische Universit\"at M\"unchen, D-85748 Garching, Germany}

\author{Safe Khan}
\affiliation{London Centre for Nanotechnology, University College London, United Kingdom}

\author{Shinichiro Seki}
\affiliation{Department of Applied Physics, The University of Tokyo, Bunkyo, Tokyo, 113-8656, Japan}

\author{Hidekazu Kurebayashi}
\affiliation{London Centre for Nanotechnology, University College London, United Kingdom}

\author{Christian H. Back}\email{christian.back@tum.de}
\affiliation{Physik-Department, Technische Universit\"at M\"unchen, D-85748 Garching, Germany}
\affiliation{Munich Center for Quantum Science and Technology (MCQST), D-80799 M\"unchen, Germany}


\begin{abstract}
Recently, it has been shown that the chiral magnetic insulator Cu$_2$OSeO$_3$ hosts skyrmions in two separated pockets in temperature and magnetic field phase space.
It has also been shown that the predominant stabilization mechanism for the low-temperature skyrmion (LTS) phase is via the crystalline anisotropy, opposed to temperature fluctuations that stabilize the well-established high-temperature skyrmion (HTS) phase.
Here, we report on a detailed study of LTS generation by field cycling, probed by GHz spin dynamics in Cu$_2$OSeO$_3$.
LTSs are populated via a field cycling protocol with the static magnetic field applied parallel to the $\langle{100}\rangle$ crystalline direction of plate and cuboid-shaped bulk crystals. By analyzing temperature-dependent broadband spectroscopy data, clear evidence of low-temperature skyrmion excitations with clockwise (CW), counterclockwise (CCW), and breathing mode (BR) character at temperatures below $T$ = 40~K are shown. 
We find that the mode intensities can be tuned with the number of field-cycles below the saturation field. By tracking the resonance frequencies, we are able to map out the field-cycle-generated LTS phase diagram, from which we conclude that the LTS phase is distinctly separated from the high-temperature counterpart. We also study the mode hybridization between the dark CW and the BR modes as a function of temperature. By using two Cu$_2$OSeO$_3$ crystals with different shapes and therefore different demagnetization factors, together with numerical calculations, we unambiguously show that the magnetocrystalline anisotropy plays a central role for the mode hybridization. 
\end{abstract}

\maketitle

\section{Introduction}
Chiral magnets with large spin-orbit interaction and significant Dzyaloshinskii-Moriya interaction (DMI) host different non-collinear spin configurations such as the helical, conical and skyrmion lattice configurations as well as the field polarized (ferrimagnetic) state in large applied magnetic fields. Many chiral magnets share a common magnetic phase diagram with a skyrmion lattice phase stabilized by thermal fluctuation located in a small pocket in temperature and magnetic field space in the vicinity of the Curie temperature~\cite{2009_Muhlbauer_Science, 2010_Yu_Nature, 2012_Seki_PhysRevB, 2012_Adams_PhysRevLett, 2012_Seki_Science, 2012_Seki_PhysRevB, 2016_Zhang_NanoLett, 2016_Qian_PhysRevB}. Furthermore, their excitation dynamics in the GHz regime has been theoretically predicted and experimentally shown in prior studies~\cite{2012_Mochizuki_PhysRevLett, 2012_Onose_PhysRevLett, 2013_Okamura_NatCommun, 2015_Schwarze_NatMater}.\\ 

The recent discovery of a novel low-temperature skyrmion (LTS) phase in various materials, including MnSi and Co$_8$ZnMn$_4$~\cite{2016_Oike_NatPhys, 2016_Karube_NatMater}, has consequently sparked renewed interest in the particular chiral magnetic insulator Cu$_2$OSeO$_3$~\cite{2018_Chacon_NatPhys,2018_Halder_PhysRevB, 2018_Qian_SciAdv, 2019_Bannenberg_npjQuantumMater}. 
It has been shown that in this material, by applying appropriate magnetic field and temperature protocols, a low-temperature skyrmion phase can be nucleated, which is distinct from the high-temperature skyrmion (HTS) phase and which seems to be stabilized by a combination of DMI and the magnetocrystalline anisotropy~\cite{2016_Gungordu_PhysRevB, 2018_Chacon_NatPhys,2018_Halder_PhysRevB}. 
As a consequence, it only appears if the externally applied magnetic field is oriented along a certain crystallographic direction.
Furthermore, in contrast to the HTS phase, the nucleation process for the LTS is nontrivial: the metastable tilted conical phase seems to be a necessary prerequisite for the formation of the LTS lattice~\cite{2018_Chacon_NatPhys,2018_Qian_SciAdv,2018_Halder_PhysRevB}. 
A recent study by Halder~\textit{et al.}~\cite{2018_Halder_PhysRevB} has shown that, to a leading order, a cubic magnetic anisotropy term is responsible for stabilizing the LTS lattice and that the characteristics of the phases can be described rather well using a Ginzburg-Landau formalism adapted for chiral magnets with cubic anisotropy~\cite{2011_Rossler_JPhysConfSer, 2016_Bauer_Book}. 
The same formalism can be extended to compute the spin excitations in both LTS and HTS lattice phases.
The description above suggests that one may influence the spatial configuration of the magnetization by controlling the strength of the magnetocrystalline anisotropy, e.g. by temperature.\\

From the above studies, it is reasonable to conclude that the stabilization mechanisms responsible for the LTS phase originate from the magnetic anisotropies when the magnetic field is applied parallel to $\langle{100}\rangle$. 
Moreover, recently, the dynamic modes of this new LTS phase have also been thoroughly investigated using low-temperature broadband ferromagnetic resonance experiments, allowing in addition the identification of the low temperature elongated skyrmion phase~\cite{2021_Back_PhysRevLett}.\\

In this paper, we address the dynamic modes in the LTS by applying broadband ferromagnetic resonance experiments as a function of temperature, $T$, and magnetic field, H.
We report the influence of the number of field cycles on the intensity of the resonant spin excitations in the LTS phase and show that a large volume fraction of the investigated spin structure in bulk crystals can indeed be converted into the LTS phase by appropriate field-cycling~\cite{2018_Chacon_NatPhys,2021_Mettus_unpublished}.
Second, we observe the low-lying breathing and counterclockwise modes and find evidence of the clockwise gyrational modes at higher frequencies~\cite{2012_Onose_PhysRevLett, 2012_Mochizuki_PhysRevLett, 2015_Schwarze_NatMater}.
Using the dynamic modes as experimental evidence for the LTS, we map out the phase diagram as a function of H and $T$.
Finally, we systematically trace the mode hybridization between a dark higher-order clockwise mode and the breathing mode~\cite{2021_Takagi_PhysRevB} as a function of temperature for two differently shaped crystals. 
We observe that the demagnetization field contribution to the gap size is negligible, consequently revealing the linear dependence on the strength of the  magnetic anisotropies. 
Our experimental results are accompanied by calculations based on a Landau-Ginzburg~\cite{2018_Chacon_NatPhys,1980_Landau_Book} energy functional, which includes the cubic magnetic anisotropy term that effectively reproduces the experimentally observed features. 
Discrepancies between the experimentally and theoretically observed size of the hybridization gap can be tentatively attributed to the influence of a gradient term of the exchange interaction.

\section{Experimental Procedures} 

\noindent{\bf Material and phase diagram.}
Copper-oxoselenite, Cu$_2$OSeO$_3$ is a chiral magnetic insulator with small Gilbert damping which crystallizes in a non-centrosymmetric cubic structure hosting 16 copper Cu$^{2+}$ ions per unit cell. The magnetization properties of the crystal are owed by the formation of tetrahedral clusters by four Cu$^{2+}$ spins in a 3-up-1-down spin configuration, behaving as a “spin-triplet” with the total spin S = 1. For the detailed visualization of the magnetic structure, see Ref.~\cite{2012_Seki_Science}. The spin-spin interactions in the system are governed by the symmetric Heisenberg interaction, which favours collinear spin structures, and an anti-symmetric Dzyaloshinskii-Moriya component analogous to $\overrightarrow{S_{\rm i}} \times \overrightarrow{S_{\rm j}}$, which make twist-like structures of neighbouring spins. Below $T_{\rm c}$ $\approx$ 58~K, the system behaves like a typical chiral magnet, hosting various magnetic phases including helical, conical, ferromagnetic and skyrmions under the influence of applied magnetic field due to a combined effect of Heisenberg and Dzyaloshinskii-Moriya (DM) interactions.\\

\begin{figure}[H]

        \centering
        \includegraphics[scale=0.55]{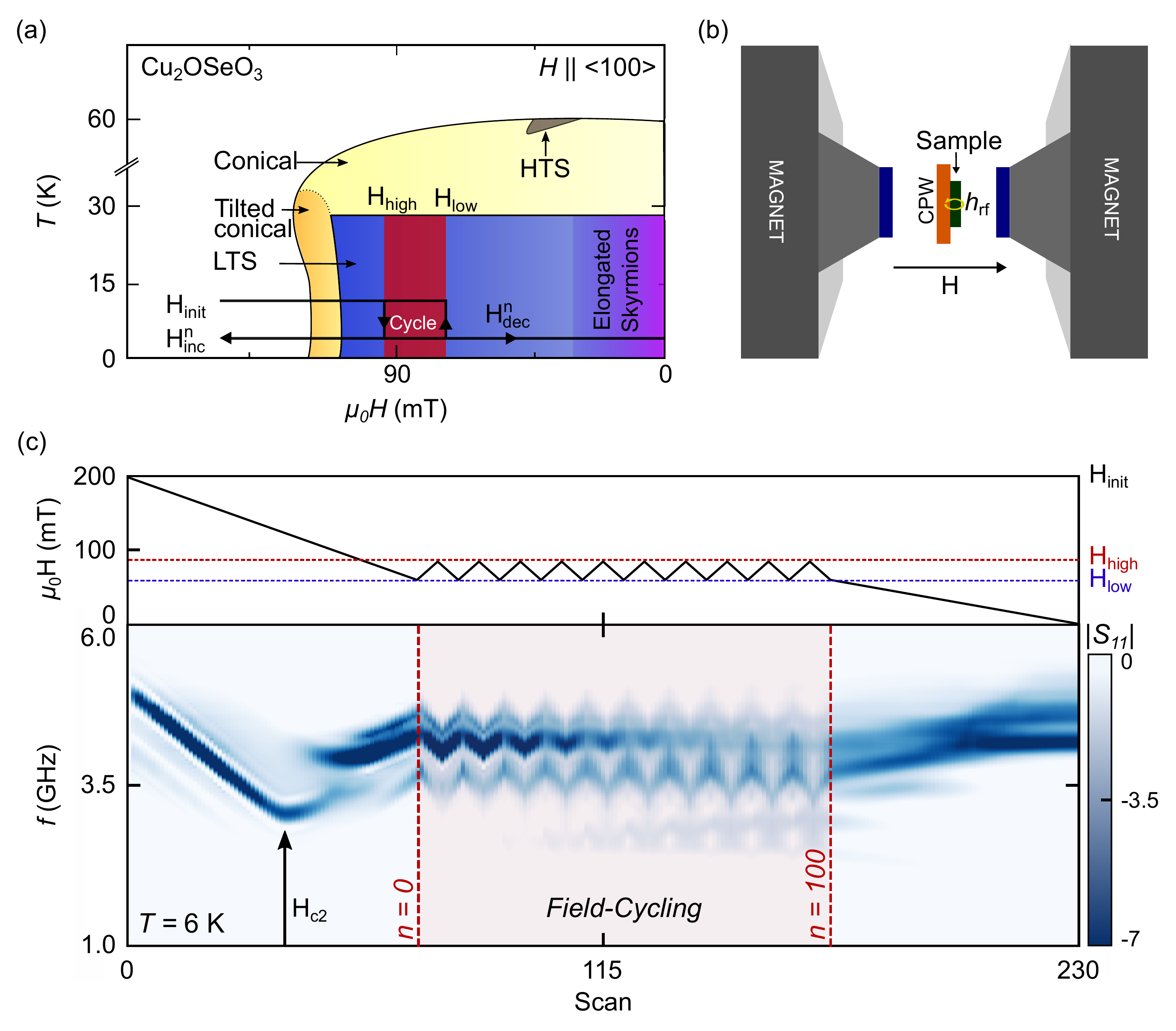}
        \caption{\textbf{Illustration of magnetization phases under field cycling and experiment schematic.} (a) Magnetic phase diagram of Cu$_2$OSeO$_3$ under field-cycling after high-field cooling. (b) Schematic representation of experimental setup (see the text for details). (c) (Top) Magnetic field as a function of scan number. (Bottom) Typical measurement profile of field-cycling protocol for $n$~=~100 cycles at 6~K. Note the color depth encodes the intensity of the reflection signal, $|S_{\rm 11}|$. The region indicated by the dashed red lines corresponds to the measurements during field-cycling, where the applied external magnetic field alternates linearly, as indicated in Fig.~\ref{fig:Fig_1}(a) and the top panel. 'Scan' indicates a single frequency sweep at a fixed magnetic field $\mu_0 H$.}
        \label{fig:Fig_1}
        
\end{figure}

To put our experimental results into context, we first show the generic magnetic phase diagram of Cu$_2$OSeO$_3$ for static magnetic fields applied parallel to the $\langle{100}\rangle$ crystallographic direction, as illustrated in~Fig.~\ref{fig:Fig_1}(a).
Above the transition temperature $T_{\rm c} \approx$ 58 K, the system is paramagnetic. 
Below $T_{\rm c}$ and without applied field, the system is described by a multidomain helical phase.
When the magnetic field exceeds a critical field value H$_{c1}$, a phase transition into the conical phase occurs.
By further increasing the magnetic field strength, the cone angle between the spins and the field axis decreases, resulting in a fully polarized state at H$_{c2}$.
In addition to these states, a topological nontrivial HTS phase exists in a small temperature and field pocket, close to $T_{\rm c}$.
This phase pocket can be extended considerably using various procedures, such as applying particular temperature-quenching techniques, applying pressure or using thin-film crystals where the thickness is of the order of several pitch lengths of the helical spiral ($\approx$ 61~nm in Cu$_2$OSeO$_3$)~\cite{2016_Rybakov_NewJPhys, 2016_Oike_NatPhys, 2016_Okamura_NatCommun, 2016_Karube_NatMater, 2017_Nakajima_SciAdv, 2019_Birch_PhysRevB, 2019_Bannenberg_PhysRevB, 2015_Wu}.
Recently, it has been demonstrated that the second, low-temperature, skyrmion phase can be nucleated after traversing the so-called tilted conical phase and following a unique magnetic field-cycling protocol~\cite{2021_Mettus_unpublished, 2021_Back_PhysRevLett}.
To enter the tilted conical phase, a state where the helix pitch tilts away from the field axis, the temperature needs to be decreased below $T\approx\mathrm{30}$~K. 
Then, as the applied field decreases further, the LTS phase emerges. Similarly, increasing the magnetic field from zero magnetic field, the tilted conical phase may form upon increasing magnetic field, again leading to the LTS generation. 
In both cases, the volume fraction of the LTS can be drastically increased by cycling the magnetic field within magnetic field boundaries where the LTS may be expected, as we will demonstrate below.\\

\noindent{\bf Experimental setup.}
We used two differently shaped bulk Cu$_2$OSeO$_3$ crystals for our experiments to vary the demagnetization fields in the samples.
One sample is a cuboid with dimensions $1.6 \times 1.6 \times 1.0~\mathrm{mm}^{3}$ with all sides orientated along $\langle{100}\rangle$.
The corresponding demagnetizing factors are 0.28, 0.28, and 0.44.
The second sample is a 0.3~mm thick plate with approximate dimensions $1.9 \times 1.4~\mathrm{mm}^{2}$ in the plane and a surface normal along the $\langle{100}\rangle$ direction.
In this case, the corresponding demagnetizing factors are $\approx$ 0.17, 0.12, and 0.71, with 0.71 being the value for the surface normal direction.
The samples were carefully polished and placed on a coplanar-waveguide~(CPW) with a $(001)$ surface facing down.
The CPW with the sample on top is introduced into a closed-cycle cryostat, inserted into the gap of an electromagnet, as illustrated in Fig.~\ref{fig:Fig_1}(b). The magnetic field is aligned normal to the plane of the CPW, i.e. along the $\langle{100}\rangle$ direction of the Cu$_2$OSeO$_3$ crystals. 
Since the samples are larger than the centre line (width 1.25~mm) of the CPW, magnetization dynamics are excited with both in-plane and out-of-plane oscillating magnetic fields, resulting in excitation of gyration and breathing modes respectively.\\

~We have employed a microwave absorption spectroscopy to probe and identify these states, as demonstrated by Y. Onose ~\textit{et al.}~\cite{2012_Onose_PhysRevLett}
The microwave magnetic fields are generated in the range of 1~-~6 GHz in steps of $\approx$~3~MHz by connecting the output port of a vector network analyzer (VNA) to the CPW. 
We measure $|S_{\rm 11}|$ or $|S_{\rm 21}|$, the reflection and the transmission signal, respectively, of the microwave absorption spectra as a function of the applied field and subtract a background signal by re-measuring the spectra in a high magnetic field well above $H_{\rm c2}$, i.e. in the field polarized state (see Supplementary Fig.~\ref{fig:Fig_S1} for details).
The measurements were carried out in a temperature range of $\mathrm{\textit{T}=5-60}$~K.\\

\noindent{\bf Field cycling protocol.} 
After high-field cooling to the desired temperature from above the ordering temperature, the applied field is first reduced from a large magnetic field H$_{\rm init}$, within the field-polarized state to the upper boundary of the desired cycling window, H$_{\rm high}$ where we subsequently start the cycling procedure. In this process, the magnetic field is repeatedly decreased and then increased between H$_{\rm high}$ and H$_{\rm low}$ in steps of $\Delta$H. 
In this context, one complete cycle means that the magnetic field is ramped down to H$_{\rm low}$ and subsequently increased again to its original value, i.e., H$_{\rm high}$ $\rightarrow$ H$_{\rm low}$ $\rightarrow$ H$_{\rm high}$. 
Upon completing $n$ number of cycles, the magnetic field is either decreased to zero-field, H$_{\rm dec}$ or increased to H$_{\rm init}$.\\

Fig.~\ref{fig:Fig_1}(c) illustrates a typical measurement profile of the field-cycling protocol for H$_{\rm init}$ = 200~mT, H$_{\rm high}$ = 80~mT, H$_{\rm low}$ = 40~mT and $\mathrm{\textit{n} = 100}$ cycles for the plate-shaped sample, recorded at 6~K, showing the normalized $|S_{\rm 11}|$. 
Note that the colour depth reflects the intensity of the detected dynamic modes.
Each scan outside the cycling region refer to the measurements of $|S_{\rm 11}|$ at particular fields, separated by steps of 1~mT. 
For example, scans 0 and 1 correspond to single measurements at H$_{\rm init}$ and H$_{\rm init}$ + 1~mT. In the cycling region, the data was recorded every 10$_{\rm th}$ cycle, thus separated by 10~mT.
Therefore, each lower vertex of the triangular-shaped signal corresponds to a completion of 10 cycles.\\

As the field decreases from H$_{\rm init}$ to H$_{\rm high}$, an almost linear slope assigned to the Kittel resonance mode in the field-polarized regime is observed.
Note that less prominent modes above and below the Kittel mode are standing spin waves across the whole Cu$_2$OSeO$_3$ crystal, which can be observed due to the low damping of the material at low temperatures~\cite{2021_Che}. 
The gradient change around $\mathrm{scan = 40}$ indicates the phase transition to the conical phase, exhibiting a frequency increase with decreasing magnetic field. 
At around scan = 60, the spectrum reveals a significant discontinuity, which we attribute to the onset of the tilted conical phase~\cite{2021_Back_PhysRevLett}. 
In the cycling regime, signatures of emergent resonant modes are observed at lower frequencies with their intensities increasing with the number of cycles.
Completing the cycling protocol, both gyration and breathing LTS modes are enhanced and clearly observed at low frequencies when further decreasing the field towards zero magnetic field.

\section{Results and Discussion}

\begin{figure}[H]

        \centering
        \includegraphics[scale=0.68]{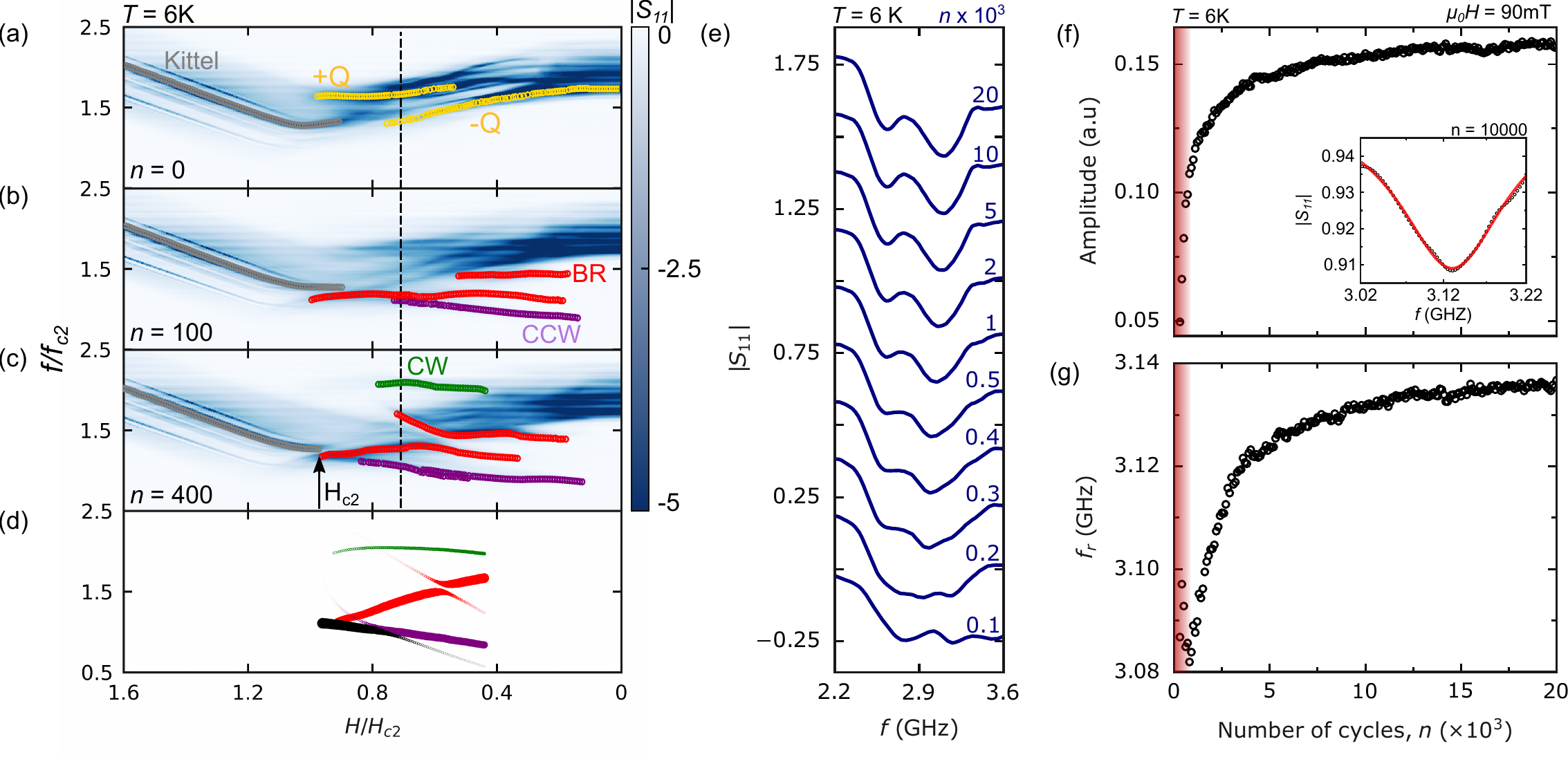}
        \caption{\textbf{Comparison of measured and calculated excitation spectra for the platelet sample and the evolution of field cycling.} (a - c) Experimental data for microwave absorption spectra recorded at 6~K for 0, 100, and 400 cycles, respectively. Fitted dots represent various magnetization phases. The dotted line is drawn at $\mu_0 H$ = 90~mT, i.e., the field value at which the plots shown in (e~-~g) are evaluated. (d) Calculated excitation spectrum with anisotropy constant $K$ = 0.0002. (e) Microwave reflection difference $|S_{\rm 11}|$ for various number of cycles $n$ at $T$ = 6~K. (f) The amplitude and the resonance frequency (g), as a function of field cycling. The results are extrapolated from the fitting routine as shown by the inset ($n$ = 10000 cycles). Note that the shaded region indicates the background signal, where the LTS lattice is slowly nucleating. For further details, see~\ref{fig:Fig_S2}.}
        \label{fig:Fig_2}

\end{figure}

Let us now discuss the magnetic excitations detected on the plate-shaped sample. Figs.~\ref{fig:Fig_2}(a~-~c) show the combined excitation spectra of H$_{\rm inc}$ and H$_{\rm dec}$, measured for a various number of cycles $\mathrm{\textit n}$ as a function of the magnetic field at $\mathrm{\textit T = 6}$ K.
In the upper panel of Fig.~\ref{fig:Fig_2}, we show data with $\mathrm{\textit{n} = 0}$, i.e. without field-cycling. 
In reminiscence of the conical state, two modes may be distinguished, exhibiting an increase in absorption frequency for decreasing magnetic field.
These resonance branches, which are expected to couple to the perpendicular excitation field with respect to the helix pitch vector, are therefore identified as $\pm$Q modes~\cite{2012_Seki_PhysRevB, 2021_Back_PhysRevLett}.
Their extracted resonance positions are marked by yellow dots.
Next, we investigate the evolution of the excitation spectrum with the number of field-cycles $\mathrm{\textit{n} = 100}$ (Fig.~\ref{fig:Fig_2}(b)) and $\mathrm{\textit{n} = 400}$ (Fig.~\ref{fig:Fig_2}(c)).
We observe the emergence of additional low lying modes, which become more pronounced with the number of cycles and for, $\mathrm{\textit{n} = 400}$ unambiguously resemble the calculated resonances shown in Fig.~\ref{fig:Fig_2}(d).
For $\mathrm{\textit{n} = 400}$, a clear signature of a hybridization gap between the BR mode and a dark mode, not showing up in the experiment, can be observed at approximately 0.5$H_{c2}$, which corresponds to $\mu_0H\mathrm{\approx75}$~mT.\\

To further highlight the evolution of the spectra under the influence of field cycling, we show in Fig.~\ref{fig:Fig_2}(e) the line-cuts at a fixed magnetic field at $\mu_0H\mathrm{ = 90}$~mT, indicated by the dashed line in Fig.~\ref{fig:Fig_2}(a~-~c) for a different number of field cycles.
On the same note, we also show the evolution of the mode amplitude and frequency $f_{r}$ in Figs.~\ref{fig:Fig_2} (f)~-~(g), respectively.
The field is chosen to be high enough in order to minimize the effects of the mode hybridization occurring around 75~mT.
We notice that after their initial regime up to 500 cycles, both parameters steeply increase with cycle number up to around 4000 number of cycles and then start to saturate afterwards. 
Since the remnants of the tilted conical phase strongly contribute to the absorption spectra for small numbers of cycles ($<$500), the fitted amplitudes and resonance positions might be inaccurate, as indicated by the shaded region.\\

\begin{figure}[H]

        \centering
        \includegraphics[scale=0.75]{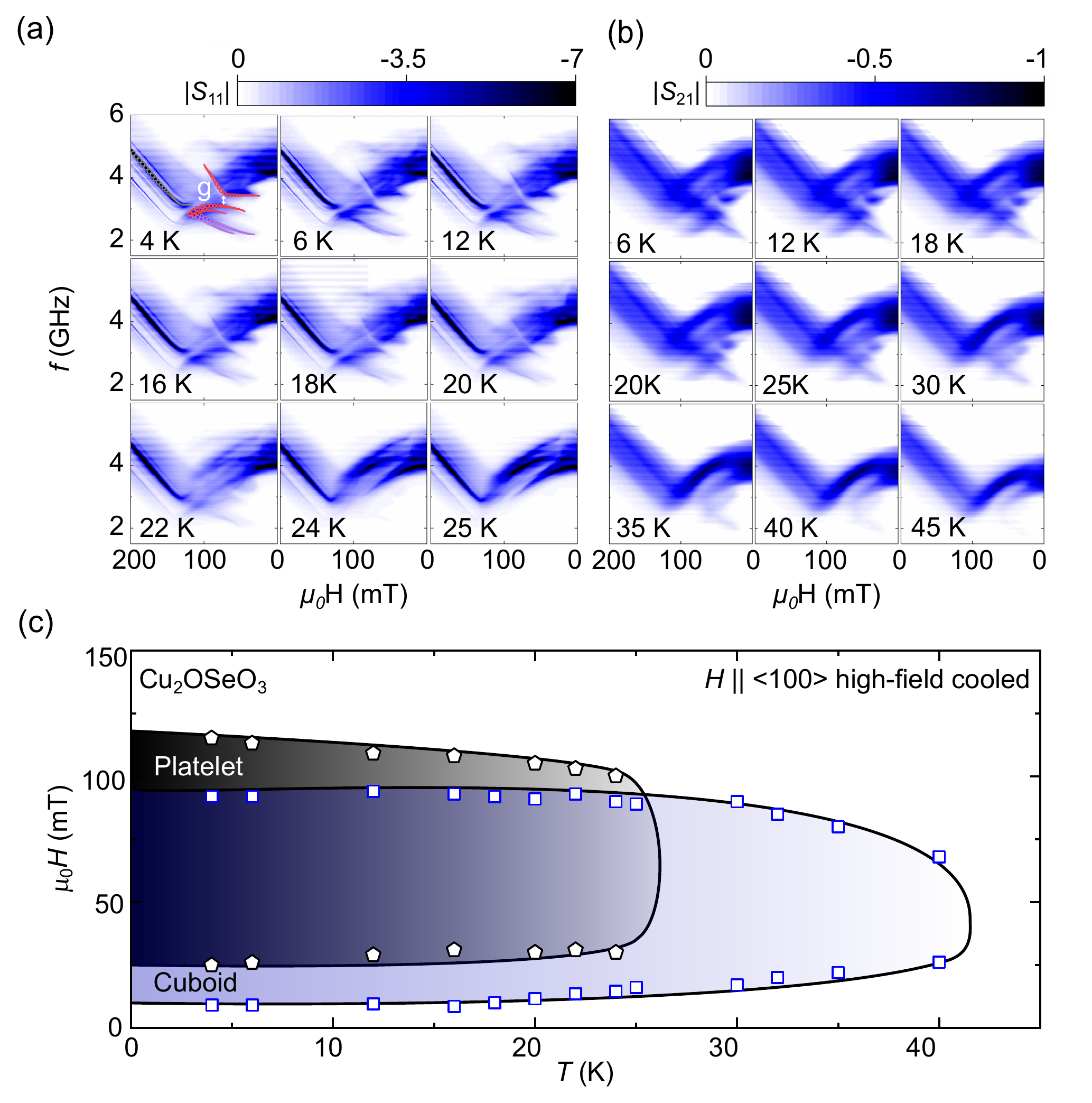}
        \caption{\textbf{Temperature evolution of microwave spectra for platelet and cuboid samples.} Collective spin excitation spectra at various temperatures and corresponding magnetic phase diagram for the platelet and cuboid-shaped samples. (a) shows the reflection difference $|S_{\rm 11}|$ for the platelet sample at various temperatures for $\mathrm{\textit{n} = 800}$. (b) Transmission difference $|S_{\rm 21}|$ for the cuboid sample for $\mathrm{\textit{n} = 1000}$ cycles. Note that a clear hybridization gap, $g$, is observed for both samples, decreasing with increasing temperature. (c) Phase diagram elaborating the LTS phase boundaries for both platelet and cuboid samples.}
        \label{fig:Fig_3}

\end{figure}

From similar measurements as a function of temperature shown in Figs.~\ref{fig:Fig_3}(a) and (b), we extract the phase pockets of the LTS phase by monitoring the respective resonances for the plate-shaped (Fig.~\ref{fig:Fig_3}(a)) and the cuboid-shaped samples (Fig.~\ref{fig:Fig_3}(b)), and present the corresponding phase diagram in Fig.~\ref{fig:Fig_3}(c).
In the plate-shaped sample, as depicted in Fig.~\ref{fig:Fig_3}(a), the LTS modes are observed below $\mathrm{\textit T = 24}$~K; however, the hybridization is no longer distinctive above $\mathrm{\textit T = 20}$~K. At $\mathrm{25}$~K, the LTS phase completely disappears while a subtle signal can still be seen at $\mathrm{24}$~K. Therefore, we extract a steep phase boundary. On the other hand, for the cuboid-shaped sample shown in Fig.~\ref{fig:Fig_3}(b), we notice that the LTS phase can be observed up to $\mathrm{\textit T = 40}$~K, although the intensities of its modes are much lower than those of the LTS phase for the plate-shaped sample.
In particular, the hybridization for the cuboid sample occurs at a lower field, at around 50~mT.\\

In order to support these results theoretically, we employ the previously established phenomenological model for chiral magnets, which can be found in various publications ~\cite{2015_Schwarze_NatMater,2018_Chacon_NatPhys, 2021_Back_PhysRevLett}.
The theoretical approach is based on the Ginzburg-Landau energy density, which consists of the following contributions:
\begin{equation}
F[\mathbf{M}] = F_0[\mathbf{M}] + r_0\mathbf{M}^2 + U\mathbf{M}^4.
\end{equation}
The first energy term $F_0[\mathbf{M}]$ represents exchange, Dzyaloshinsky-Moriya and dipolar interactions, characterized by their respective strengths, $J$, $D$ and $\tau$.
The shape of the samples under investigation enters the latter by the corresponding demagnetization tensor $N$ with tr($N$) = 1.
Additionally, the Zeeman term, with magnetic field $\mathbf{B}$ and the cubic anisotropy, with anisotropy strength $K$ are included in $F_0[\mathbf{M}]$. 
The remaining terms are governed by the Ginzburg-Landau coefficients $r_0$ and $U$.
While $U$ is assumed to be constant and greater than zero, $r_0$ is set to be the measure of the distance to the critical temperature $T_\text{c}$ up to  the first order and therefore $r_0 \sim T - T_\text{c}$. Rescaling the theory reduces the number of parameters, leaving $r_0$, $N$, $K$ and $\mathbf{B}$ as variables. The dipolar interaction strength was already determined in prior reports and is set to $\tau \approx 0.88$ accordingly \cite{2015_Schwarze_NatMater,2018_Chacon_NatPhys}.\\

Since the focus of this work lies in the investigation of the LTS phase, we likewise limit our analysis to the eigenmodes and eigenresonances of the topologically nontrivial skyrmion states.
In the following, the calculations are performed either for a spherical sample with demagnetization factors $N_\text{i} = 1/3$ representing the cuboid, or for a platelet sample with $N_x = 0.17$, $N_y = 0.12$ and $N_z = 0.71$. The temperature is set to $r_0$ = -1000, and the anisotropy constant to K = 0.0002, given in dimensionless units, if not stated otherwise.
Our calculations revealed that the shape of the sample under investigation does not affect the gap size significantly.\\

In Fig.~\ref{fig:Fig_2}(d), we show the calculated excitation spectrum  normalized to $H_{\rm c2}$.
Calculations are performed for the plate-shaped sample, with the magnetic field applied along the surface normal, i.e. parallel to the $\langle{100}\rangle$ direction.
The field boundaries were chosen to cover the range in which the skyrmion phase is energetically more favorable than the topological trivial states. 
Nevertheless, it was already shown that the LTS phase may remain even in smaller fields, merging into the elongated skyrmion phase, see Ref.~\cite{2021_Back_PhysRevLett}. 
The LTS phase is characterized by three distinct modes of the skyrmion lattice, the counterclockwise (CCW) gyrational mode, which decreases in frequency with decreasing magnetic field (violet), the breathing (BR) mode, which increases in frequency with decreasing magnetic field (red) 
and the clockwise (CW) gyrational mode again with decreasing frequency as a function of decreasing magnetic field (green). We find excellent agreement between the calculated excitation spectrum and the experimentally detected dynamic modes after field cycling.\\ 

\begin{figure}

       \centering
        \includegraphics[scale=0.28]{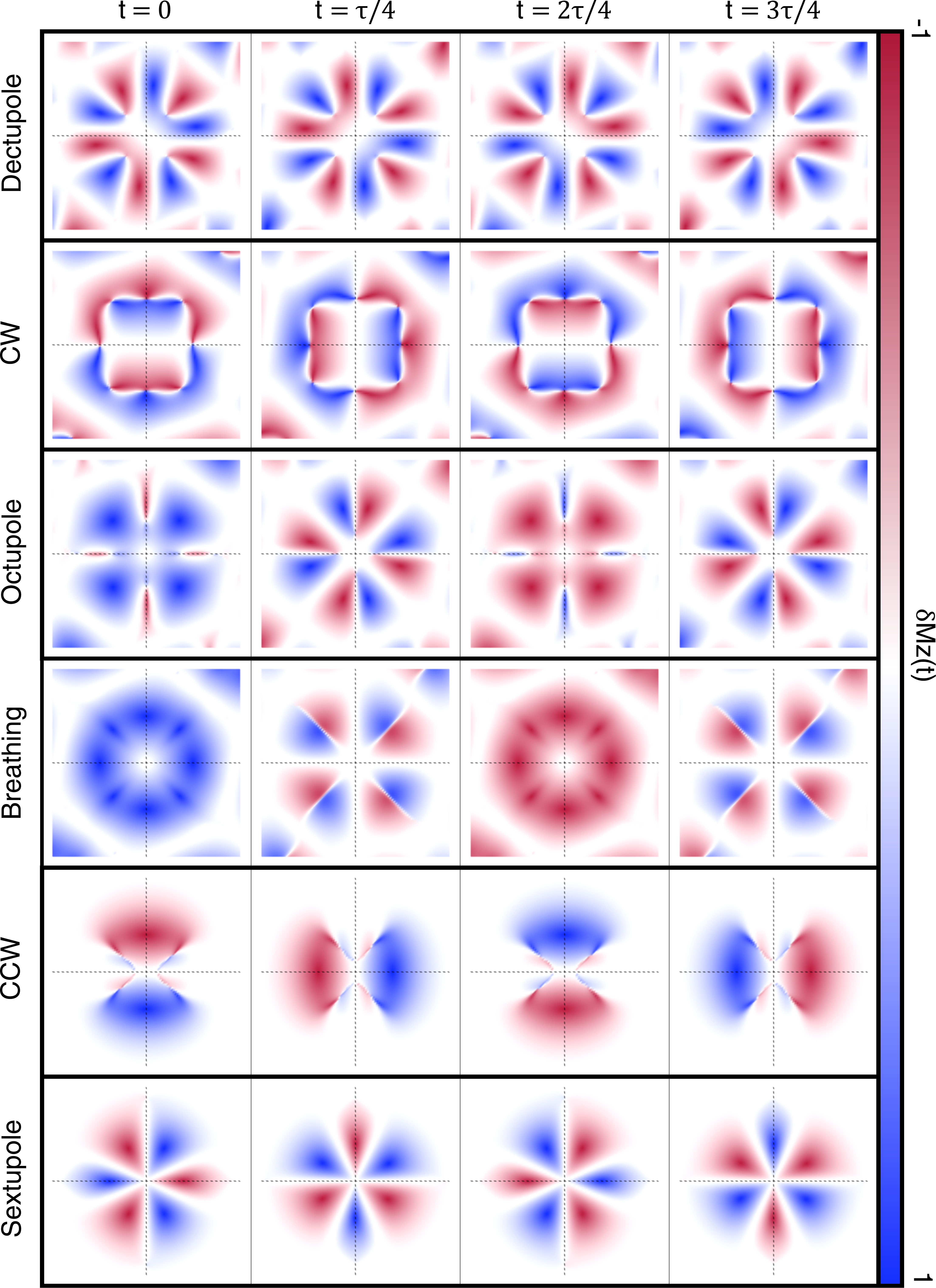}
        \caption{\textbf{Calculated real-space images of skyrmions under resonant excitation of the six lowest-frequency modes for K = 0.0002.} The columns correspond to different characteristic moments during the oscillation; notably, they are separated in time by one-fourth of the period $\tau$. For clarity, only the normalized dynamic magnetization component $\delta$$M_z$(t) is shown here. From top to bottom, the modes are identified as the dectupole, clockwise gyration mode (CW), octupole, the breathing mode (BR), counterclockwise (CCW) and the sextupole gyration.}
        \label{fig:Fig_4}
      
\end{figure}

A particular feature that becomes very clear from the theoretical analysis is the opening of a gap for the BR mode (red) which hints to a hybridization and which is also observed in the experiments. This hybridization feature is not observed in the HTS phase and will be discussed in the following. 
By introducing the cubic anisotropy term in our micromagnetic calculations, we observe a hybridization of the breathing, clockwise and counterclockwise mode with additional dark skyrmionic modes, i.e. modes that cannot be observed in the experiment due to low spectral weight. These anti-crossings differ significantly in magnetic field position, spectral weight and gap size and the dominant hybridization observed also in our experiments is the one of the BR mode. 
By means of real space images for different times $t$ (see Supplementary Fig.~\ref{fig:Fig_S3} for details), those dark modes are identified as higher-order clockwise modes. Based on the number of nodes, we refer to these modes as Sextupole (6 nodes), Octupole (8 nodes) and Dectupole (10 nodes) CW mode.
The following hybridizations are observed in our theoretical model: breathing mode - Octupole CW mode, CW mode - Dectupole CW mode, CCW mode - Sextupole CW mode.
The effect on the mode characteristic is shown in the time evolution of the dynamic magnetization component in Fig.~\ref{fig:Fig_4}.
From this, it can clearly be seen that the individual modes additionally exhibit the excitation features of their hybridization counterpart.\\

Even though more resonance branches are crossing each other, only the modes listed above lead to a repulsion due to the anisotropy.
To verify these observations, we calculate the inner product, denoted as $\bra{\textbf{\textit{v}}^\alpha} \ket{\textbf{\textit{v}}^\beta}$, of the normalized eigenstates $\textbf{\textit{v}}^\alpha$ in momentum space at different magnetic fields for K = 0 and K $\neq$ 0, respectively. Here, $\alpha$ and $\beta$ indicate the eigenvalue indices. The results are shown in Tab.~\ref{tab:Tab_1}.
The field values are chosen to be close to the centre of the observed hybridizations, in the order: CW – Dectupole CW, BR – Octupole CW, CCW - Sextupole CW.
From the calculations, it is evident that the anisotropy does not affect the orthogonality
between the eigenstates CW – Octupole CW, BR – CCW, and BR – Sextupole CW.
As linearly independent eigenstates, the crossing does not result in repulsion and hybridization, respectively.
For the remaining inner products, the value changes significantly by introducing the additional energy term.
Approaching the respective intersection point increases the inner product further, indicating a stronger interaction between the resonance modes.\\

\begin{table}
\caption{Inner product calculated for various fields, with and without cubic anisotropy.}
\centering
	\begin{tabular}{|l|l|l|l|l|l|l|}
		\hline
	&\multicolumn{2}{|c|}{$H$ = 0.42\,$H_\text{c2,0}$}& \multicolumn{2}{|c|}{$H$ = 0.63\,$H_\text{c2,0}$}& \multicolumn{2}{|c|}{$H$ = 0.83\,$H_\text{c2,0}$}\\
		\hline
	$\bra{\textbf{\textit{v}}^\alpha} \ket{\textbf{\textit{v}}^\beta}$&$K$ = 0&$K$ = 0.0002&$K$ = 0&$K$ = 0.0002&$K$ = 0&$K$ = 0.0002\\
		\hline
		CW-Dectupole&0&0.015&0&0.010&0&0.002\\
		\hline
		CW-Octupole&0&0&0&0&0&0\\
		\hline
		BR-Octupole&0&0.040&0&0.117&0&0.047\\
		\hline
		BR-CCW&0&0&0&0&0&0\\
		\hline
		BR-Sextupole&0&0&0&0&0&0\\
		\hline
		CCW-Sextupole&0&0.026&0&0.036&0&0.037\\
		\hline
	\end{tabular}
	\label{tab:Tab_1}
\end{table}

In order to approach these findings from a phenomenological point of view, we further examine the real space images of the discussed resonance modes for K = 0, regarding their symmetry characteristics.
From the visualization, we extract the number of symmetry axis $m$ for each mode, summarized in Tab.~\ref{tab:Tab_2}.
While the CW and CCW modes only exhibit one symmetry axis, the number $m$ increases for the higher-order CW modes, in accordance with their node number.
Due to the axial symmetry of the breathing mode, we denote $m$ to be infinity in this case.
Considering the hybridizing modes and their corresponding numbers of symmetry axis $m$, it follows that the cubic anisotropy imposes a symmetry selection rule, allowing hybridizations only between states with axial symmetry and even $m$, or with even or odd number $m$, respectively.\\

\begin{table}
\caption{Number of symmetry axis $m$ assigned to the respective modes.}
\centering
	\begin{tabular}{|l|c|}
		\hline
	    Mode&Number of symmetry axis $m$\\
	    		\hline
		CCW&1\\
		\hline
		CW&1\\
		\hline
		Sextupole CW&3\\
		\hline
		Octupole CW&4\\
		\hline
		Dectupole CW&5\\
		\hline
		BR&$\infty$\\
		\hline
	\end{tabular}
	\label{tab:Tab_2}
\end{table}

\begin{figure}

       \centering
        \includegraphics[width = 0.8 \textwidth]{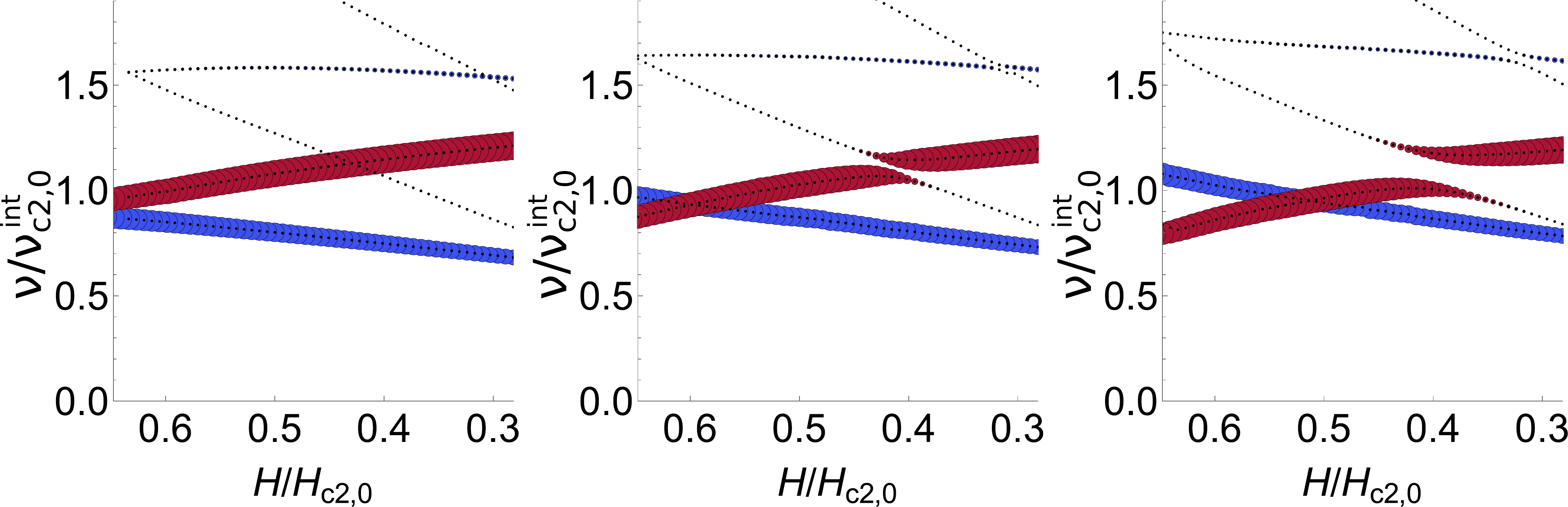}
       \caption{\textbf{Field evolution of the resonance frequencies.} Resonance frequencies as a function of the applied magnetic field for increasing anisotropy strength. From left to right: K = 0, K = 0.0002, K = 0.0004. Frequencies are given in units of the internal critical field $\nu_\text{c2,0}^\text{int} = \frac{g\mu_0\mu_\text{B}}{2\pi \hbar} H_\text{c2,0}^\text{int}$ (critical internal field $H_\text{c2,0}^\text{int}$ for K = 0) and fields in units of transition field $ H_\text{c2,0}$. The size of the dots represents the estimated spectral weight of the modes.}
        \label{fig:Fig_5}

\end{figure}

Next, we are able to account for the size of the hybridization gap $g$ quantitatively by extracting the minimal frequency difference between the breathing and octupole mode branches from the theoretical and experimental results.
The calculated resonance frequencies, normalized by
\begin{equation}
	\nu_\text{c2,0}^\text{int} = \frac{g\mu_0\mu_\text{B}}{2\pi \hbar} H_\text{c2,0}^\text{int}
\end{equation} 
are given in dimensionless units, with internal critical field $H_\text{c2,0}^\text{int}$ determined without cubic anisotropy.
For small $K$, we assume $H_\text{c2,0}^\text{int} \approx H_\text{c2}^\text{int}$, in order to obtain an estimate for the physical units.
The internal critical field is connected to the saturation magnetization $M_\text{s}$ by the internal susceptibility $\chi_\text{con}^\text{int}$ in the conical phase \cite{2015_Schwarze_NatMater,2017_Waizner_PhD_2,2017_Garst_JPhysD},
\begin{equation}
	M_\text{s} = H_\text{c2,0}^\text{int} \chi_\text{con}^\text{int}.
\end{equation}
In Fig.~\ref{fig:Fig_5}, the calculated field dependence of the excitation frequencies (black dots) is shown for K = 0, K = 0.0002 and K = 0.0004.
While the colors of the dots indicate the excitation geometry with respect to the applied magnetic field direction (blue: in-plane, red: out-of-plane), the size of the dots reflects the spectral weight of the resonance modes.
As already anticipated before, it can be seen, that including the cubic anisotropy term into the model results in hybridizations of different modes, which interaction strengths depend on the anisotropy value.\\

In order to extract the temperature dependence from our calculations quantitatively, it is sufficient to leave only the anisotropy value as a free parameter.
The Ginzburg-Landau coefficient $r_0$, the measure for the temperature, enters our model as a scaling factor proportional to the saturation magnetization $M_\text{s}$.
Therefore, by extracting the temperature-dependent parameters $H_\text{c2,0}^\text{int}$ and $M_\text{s}$, the calculated spectra can be given in physical units at the corresponding temperature.\\

The anisotropy constant $\tilde{K}$ is determined with the help of the results obtained in \cite{2018_Chacon_NatPhys,2018_Halder_PhysRevB}.
It was found that, exceeding $\tilde{K}_\text{c}^\text{th}$ $\approx 0.0001$, the LTS phase would stabilize in the theoretical model. Note that in order to avoid confusion, the dimensionless anisotropy constant is marked by a tilde, here. This value corresponds to the dimensionless value obtained from the experiments,        
\begin{equation}
\tilde{K}_\text{c}^\text{exp} = \frac{K_{\sigma,\text{c}}}{\mu_0 H_\text{c2,0}^\text{int} M_s} \approx 0.07
\end{equation}
with the anisotropy constant $K_{\sigma,\text{c}}$ given in units of energy density \cite{2018_Chacon_NatPhys}.
Considering now the observed anisotropy values in \cite{2018_Halder_PhysRevB}, the theoretical model can provide an estimate of the gap size, as a function of temperature.\\

\begin{figure}[H]

        \centering
        \includegraphics[scale=0.55]{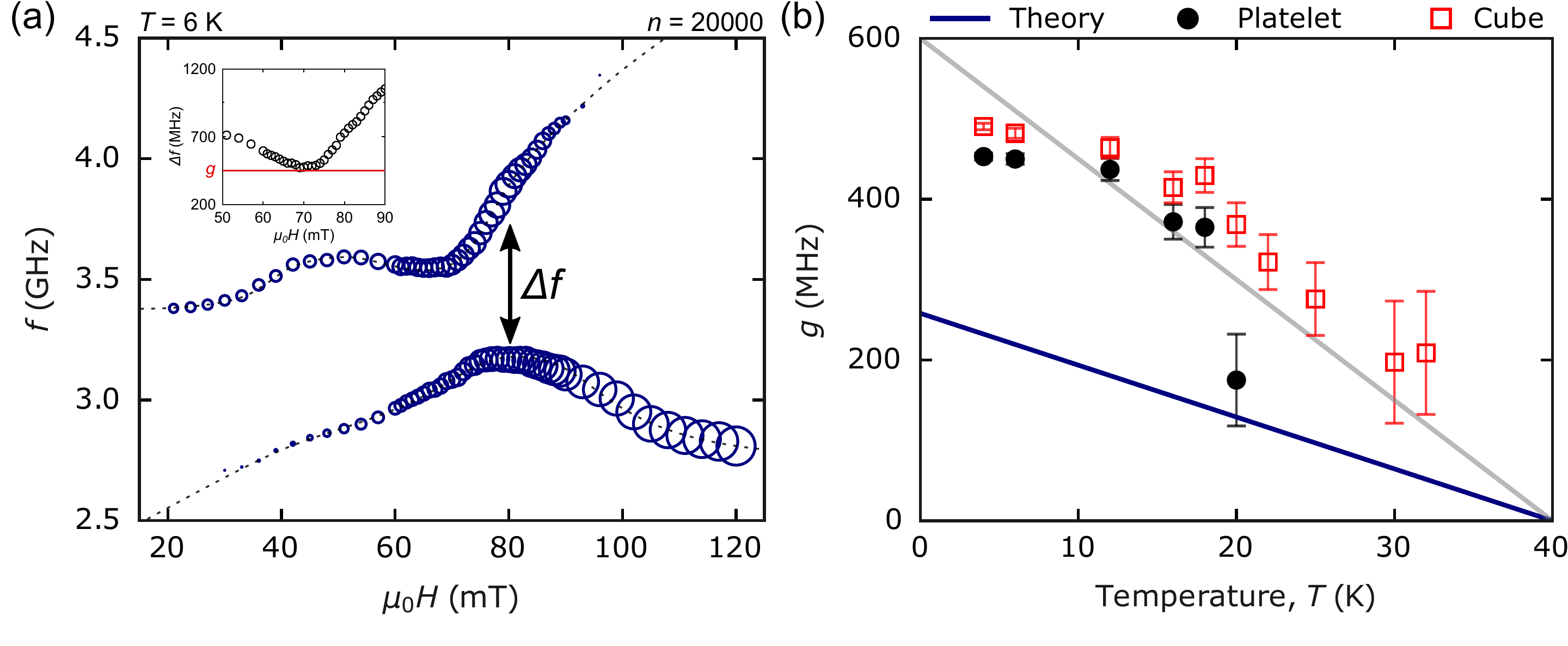}
        \caption{\textbf{Field-cycling and temperature evolution of the hybridization gap.} (a) Fitted frequency dependence of upper and lower Br modes as a function of field. The inset shows the frequency difference of the two modes, $\Delta f$, plotted as a function of field for the platelet sample at $T$ = 6~K. (b) Temperature evolution of hybridization gap, $g$, for the platelet and cuboid samples. A smooth grey line is added as a guide to the eye.}
        \label{fig:Fig_6}

\end{figure}

The experimental data is evaluated the same way as described above. The resonance positions of the breathing and octupole mode in the vicinity of the hybridization were extracted from two Lorentzian peak fits. 
Calculating the frequency difference $\Delta f$ between both branches as a function of field, we define the minimum value as the gap size $g$. 
In Fig.~\ref{fig:Fig_6}(a) the fitting results are shown for the measurements performed at 6~K. At this temperature, $g$ obtained from $\Delta f$, given in the inset, reads 450~MHz.\\

\begin{figure}[H]

        \centering
        \includegraphics[width = 0.9\textwidth]{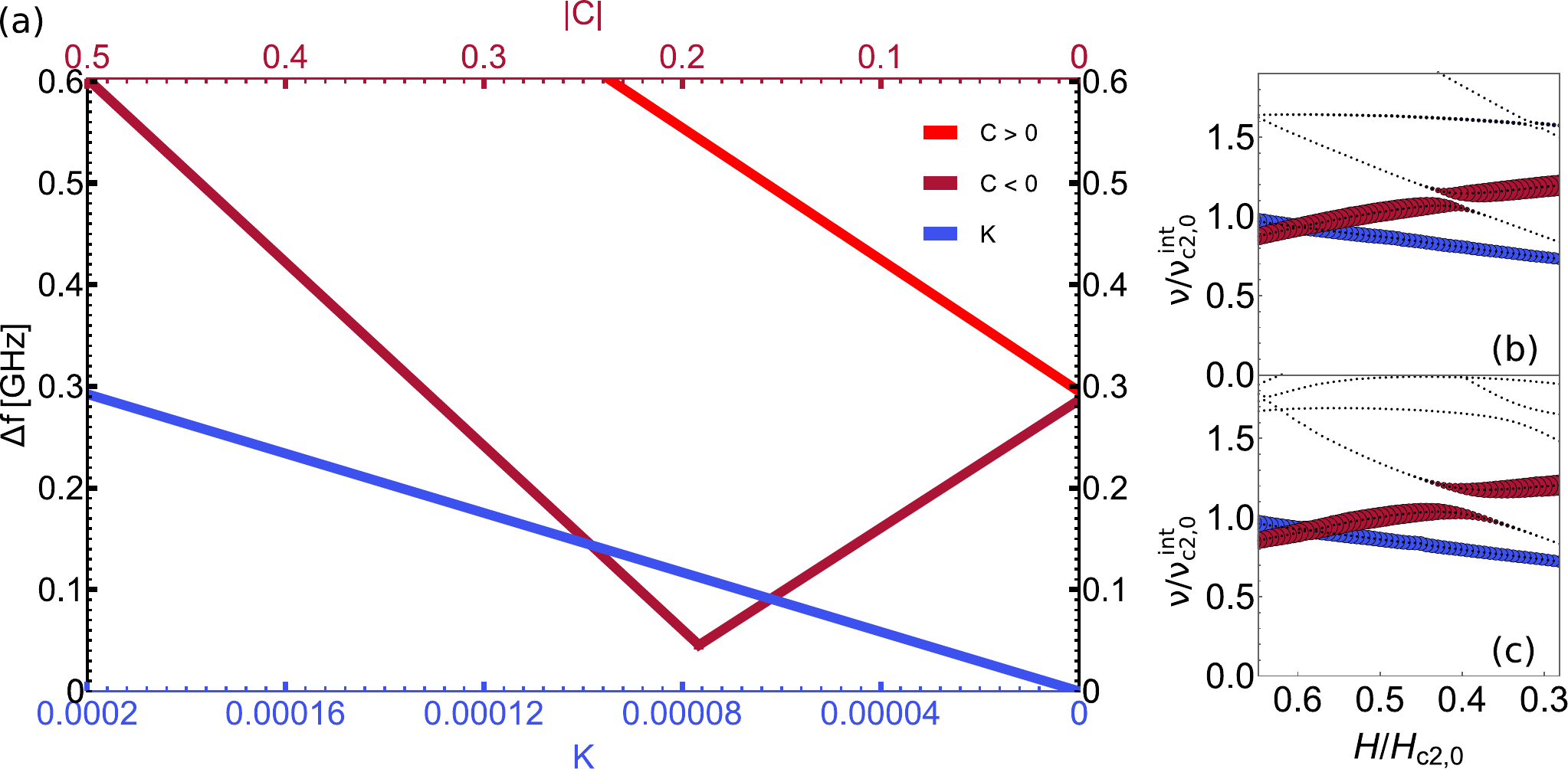}
        \caption{\textbf{Comparison of the effect of different anisotropy terms on the hybridization gap size.} (a) Theoretically expected gap size as a function of anisotropy strength. The lines depict the results, calculated with (light and dark red) and without (blue) exchange anisotropy. In the two cases (red lines), $K$ is set to be a constant  ($K = 0.0002$). (b)~-~(c) Resonance frequencies spectra as a function of the external field, for K~=~0.0002, C~=~0 and K~=~0.0002, C~=~0.2.}
        \label{fig:Fig_7}

\end{figure}

For a conclusive temperature dependence of the gap size, we extend our analysis to the temperatures covered by our experiments. Our findings from both measurements and calculations are collated in Fig.~\ref{fig:Fig_6}(b).
In both crystals studied, we observe a similar trend qualitatively, namely a decrease of $g$ with increasing temperature. 
In order to serve as a guide to the eye, a grey line is added to the figure.
Its intersection point is chosen to be consistent with the temperature-dependent anisotropy measurements from Halder~\textit{et al.}~\cite{2018_Halder_PhysRevB}.
The theoretically estimated gap size, converted to physical units, is given by the blue line, reminiscent of the behaviour in the experimental data.
Its slope, however, differs by approximately a factor of 2.3 with respect to the experiment.
A small deviation between experimental and theoretical data is induced by the conversion from dimensionless to physical units.
However, this error is assumed to be small compared to the factor mentioned above, leaving the origin of this discrepancy open.
In order to further investigate these findings, we included an additional exchange anisotropy term
\begin{equation}
F_e[\mathbf{M}] = C \sum_{\mathbf{k}} \sum_{\alpha} k_{\alpha}^2 M_\mathbf{k}^\alpha M_{-\mathbf{k}}^\alpha
\label{eqn: CAnisotropy}
\end{equation}
into the free energy functional. 
This anisotropy is allowed by the symmetry of the $P2_13$ space group, and was already taken into consideration in \cite{2019_Bannenberg_npjQuantumMater, 1980_Bak_JPhysC}.
Depending on the sign of $C$, the helix pitch aligns either along $\langle111\rangle \,(C < 0)$ or $\langle100\rangle \,(C > 0)$.
Fixing all parameters apart from $C$, we calculate the resonance spectra and therefore also the gap size, in the same manner as before, as a function of the anisotropy strength.
The extracted values are summarized in (Fig.~\ref{fig:Fig_7}).
The blue line corresponds to the results already shown in Fig.~\ref{fig:Fig_6}(b), only including the cubic anisotropy.
The red curves (light red $C > 0$, dark red $C < 0$) are obtained for both cubic and exchange anisotropy, while the cubic anisotropy parameter was set to $K = 0.0002$.
For certain exchange anisotropy strengths, independent of the sign of $C$, the gap reaches the experimentally observed value of approximately 0.6\,GHz.
In the case of positive $C$, the value increases with increasing anisotropy strength.
For negative $C$, the exchange anisotropy is counteracting the cubic anisotropy, leading first to a reduction of the gap size.
Exceeding a certain value, the exchange anisotropy is prevailing, causing the gap size to increase again.
Due to energetically more favorable alignment either along the $\langle111\rangle$ or $\langle100\rangle$ directions, the exchange anisotropy induces a distortion of the skyrmion lattice into an oblique and square lattice, depending on the external field.
Similar observation, induced by the cubic anisotropy in this case, were already reported in~\cite{2018_Chacon_NatPhys,2021_Back_PhysRevLett}.
Nevertheless, despite this additional energy contribution, the resonance spectra are left essentially unchanged, as shown in Fig.~\ref{fig:Fig_7}(b)~-~(c).
The calculations were performed for Fig.~\ref{fig:Fig_7}(b) K~=~0.0002, C~=~0 and Fig.~\ref{fig:Fig_7}(c) K~=~0.0002, C~=~0.2, respectively, illustrating the significant impact on the hybridization process.
These results suggest that further anisotropy terms~\cite{2016_Bauer_Book, 2021_Back_PhysRevLett}, like the one discussed above, might play a non-negligible role and need to be considered for a comprehensive analysis of the temperature dependence of the hybridization.\\


\FloatBarrier

\section{Summary}
In conclusion, we report a comprehensive study of the nucleation of the low-temperature skyrmion lattice that is stabilized by the magnetocrystalline anisotropy of the sample via a field-cycling protocol and systematically trace the characteristic hybridization in the breathing mode. By employing two different sample geometries, we were able to remove contributions from the demagnetizing field and investigate the dynamics of low-temperature skyrmion modes for various temperatures. Our analysis of the hybridization gap highlights that the cubic magnetocrystalline anisotropy terms govern the origin of the hybridization following a particular symmetry selection rule.
In addition, we show that in the non-collinear skyrmion phase, the contribution of the exchange anisotropy to the excitation spectra can possibly be traced via the observation of the hybridization gap and detailed theoretical analysis. Note that this contribution can not be detected in collinear phases. 


\section{Acknowledgement}
We would like to acknowledge fruitful discussions with M. Garst. O. L. thanks the Leverhulme Trust for financial support via RPG-2016-391. We further acknowledge JSPS Grants-In-Aid for Scientific Research (Grant Nos. 18H03685, 20H00349, 21H04440), JST PRESTO (Grant No. JPMJPR18L5) and Asahi Glass Foundation for funding.
This work has also been funded by the Deutsche Forschungsgemeinschaft (DFG, German Research Foundation) under SPP2137 Skyrmionics, TRR80 (From Electronic Correlations to Functionality, Project No.\ 107745057, Project G9), and the excellence cluster MCQST under Germany's Excellence Strategy EXC-2111 (Project No.\ 390814868).

\clearpage
\bibliography{Ref}

\newpage
\section{Supplementary}
\beginsupplement

\section{S.1 Experimental setup and data processing}

The magnetization dynamics of the low-temperature skyrmion phase can be determined by measuring the subtle microwave response of the magnetic system as a function of frequency and field. Thus, conventional magnetic spectroscopy techniques such as electron spin resonance (ESR) spectrometer cannot be employed. In ESR, the magnetic response to a system is probed by scanning the magnetic field at a fixed frequency. For this reason, we have performed our measurements utilizing a broadband ferromagnetic resonance (BB-FMR) technique with a vector network analyzer (VNA). The microwave line from the VNA is linked to the coplanar-waveguide (CPW) via a semi-rigid cable. Then, the RF current, $I\cos(\omega t)$ is flown through the CPW which induces a local magnetic field and directly couples with the magnetization, \textbf{M} in the sample. As the static magnetic field \textbf{\textit{H}} is varied, frequencies are swept at a fixed microwave power (0~dBm). Therefore, the microwave absorption in this reference is the change in reflection (transmission) power of the CPW that induces the driving field ($H_{\rm rf}$). The obtained spectra are denoted respectively as $S_{\rm 11}$ and $S_{\rm 21}$ for the response to reflection or transmission.\\

Fig.~\ref{fig:Fig_S1} (a) shows typical reflection spectra $S_{\rm 11}$ obtained directly from the VNA for the plate-shaped sample at $\mathrm{\textit T = 6}$~K. For systematic measurement analysis and accurate determination of excitation dynamics, the data was processed over three distinctive stages. Firstly, we carefully remove the excessive noise in the signal utilizing an algorithm involving a numerical filter. In order to preserve all necessary information from the data obtained, a lossless analytical Savitzky-Golay filter~\cite{1964_Savitzky_AnalChem} was employed, which the filter recalculates the signal from the convolution coefficients by fitting successive data points by a polynomial function using linear least squares.
The filtered signal, $S_{\rm 11}'$ is shown in Fig.~\ref{fig:Fig_S1}(b). Secondly, the spectra in the region of interest were subtracted by a background signal ($\mu_0 H$ = 300~mT) obtained at a high magnetic field, as shown in Fig.~\ref{fig:Fig_S1}(c).
Lastly, Fig.~\ref{fig:Fig_S1}(d) shows the resulting signal after a full conversion into a linear scale, i.e., $|S_{\rm 11}| = 10^{\Delta S_{\rm 11}/10}$. The data were then fitted using multi-peak Lorentzian to extract the parameters of interest. We have used the same systematic protocol for all data presented in this paper.

\setcounter{figure}{0}  

\section{S.2 Additional cycling data}

\begin{figure}[H]

        \centering
        \includegraphics[scale=0.55]{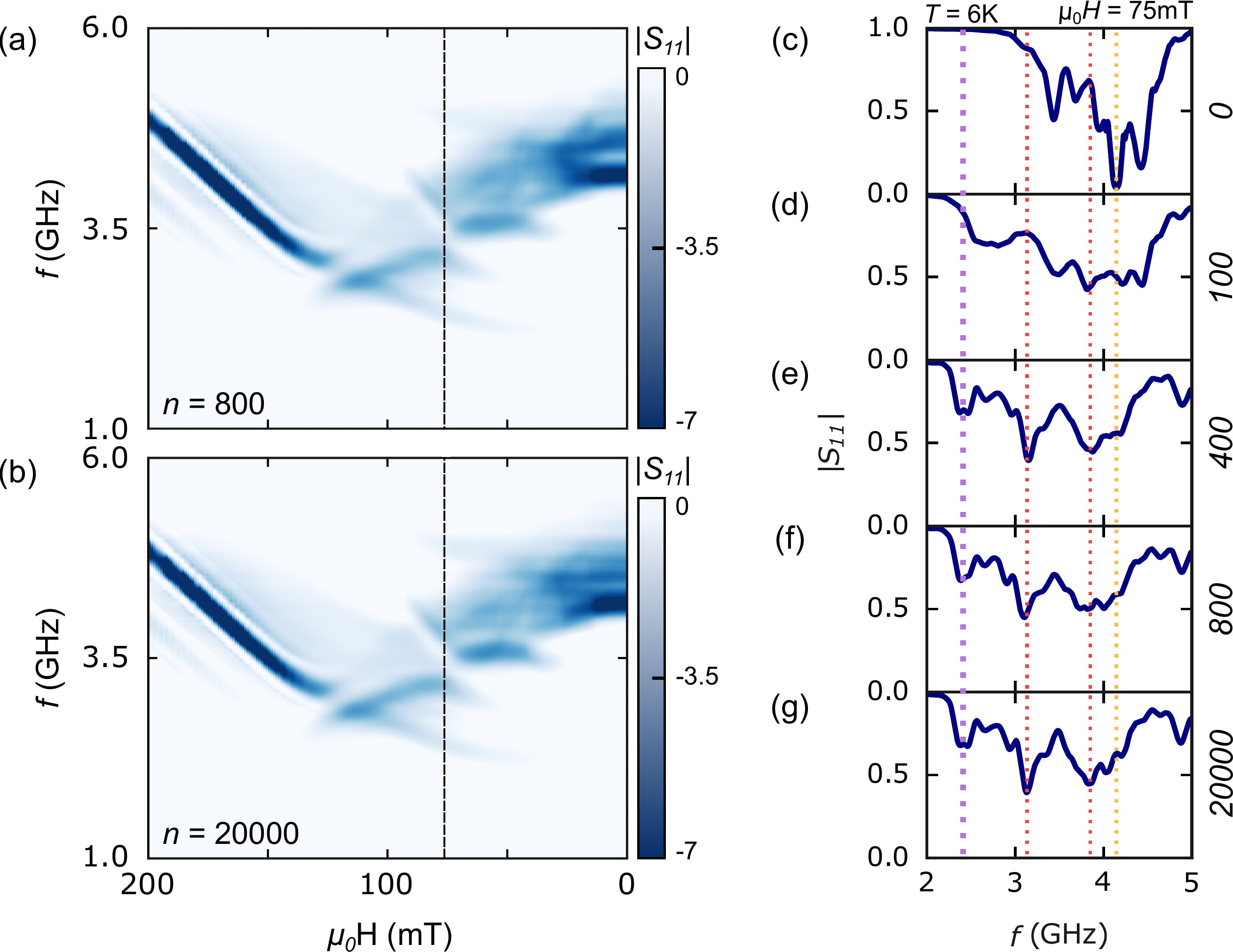}
        \caption{\textbf{Evolution of measured excitation spectra for the platelet sample.} (a~-~b) Experimental data for normalized microwave absorption spectra, $|S_{\rm 11}|$, recorded at 6~K for 800 and 20000 number of cycles, respectively. (c~-~g) $|S_{\rm 11}|$ as a function of frequency measured at $\mu_0 H$ = 75~mT (dotted vertical line in (a~-~b)) for different number of field cycles $n$, upon completing the full cycling protocol. Above 100 cycles, various dynamic modes, including CCW skyrmions (purple), Br skyrmions (red), and conical (orange), are clearly observed.}
        \label{fig:Fig_S2}

\end{figure}

Here, we present additional spectra data obtained respectively, at 800 and 20000 field cycles as shown in Figs.~\ref{fig:Fig_S2}(a~-~b) obtained at $\mathrm{\textit T = 6}$~K.
A vertical line is drawn to indicate the approximate field ($\mu_0H$~$\approx$~75~mT) where hybridization is clearly seen for both number of cycles.
In order to investigate the evolution of the dynamic modes, we further plot the line-cut data at this field for different numbers of field cycles in Fig.~\ref{fig:Fig_S2}(c~-~g).\\

Our results highlight that the intensity of the hybridization becomes more visible with the number of accumulated field cycles which must be connected to the volume fraction of the LTS phase. For example, a direct comparison between $\mathrm{\textit{n} = 100}$ and $\mathrm{\textit{n} = 400}$ illustrates clearer evidence of the hybridization gap in the LTS phase. Note that for $\mathrm{\textit{n} = 400}$, the spectrum is dominated by the excitations of the LTS phase in the field range of $\mu_0H\approx$ $\mathrm{25-125}$~mT and only little weight is observed in the $\pm$Q modes of the conical phase. Furthermore, the intensities of the LTS modes remain almost constant from $\mathrm{\textit{n} = 400}$ onwards even when the number of cycles is extended to the extreme case of $\mathrm{\textit{n} = 20000}$.

\section{S.3 Theoretical model}

The theoretical analysis presented in this report draws upon the well established Ginzburg-Landau theory for chiral magnets  ~\cite{2015_Schwarze_NatMater,2018_Chacon_NatPhys, 2021_Back_PhysRevLett}.
The energy functional $F$ consists of the following energy terms,
\begin{equation}	
\begin{aligned}
F[\mathbf{M}] =   \sum_{\mathbf{k}}
\Bigg(
&r_0 \mathbf{M}_\mathbf{k}\cdot\mathbf{M}_{-\mathbf{k}} + \frac{J}{2} (\mathbf{k} \cdot \mathbf{k} )(\mathbf{M}_\mathbf{k}\cdot \mathbf{M}_{-\mathbf{kg}}) +
i D \mathbf{M}_{-\mathbf{k}} \cdot(\mathbf{k} \times \mathbf{M}_\mathbf{k})\\
v&+ U \sum_{\mathbf{k}_2,\mathbf{k}_3,\mathbf{k}_4} (\mathbf{M}_\mathbf{k}\cdot\mathbf{M}_{\mathbf{k{}}_2}) (\mathbf{M}_{\mathbf{k{}}_3}\cdot\mathbf{M}_{\mathbf{k{}}_4}) \delta_{\mathbf{k} + \mathbf{k}_2 + \mathbf{k}_3 + \mathbf{k}_4,0} -
\mathbf{B}\cdot\mathbf{M}_0\\					
&- \tau \mathbf{M}_0 N \mathbf{M}_0 + \sum_{\mathbf{k}} \frac{(\mathbf{k}\cdot\mathbf{M}_\mathbf{k})(\mathbf{k}\cdot \mathbf{M}_\mathbf{k})}{ \mathbf{k}\cdot\mathbf{k}}\\
& - K \sum_{\mathbf{k},\mathbf{k_2},\mathbf{k_3},\mathbf{k_4}} \sum_{\alpha} \mathbf{M}_{k}^{\alpha}\mathbf{M}_{k_2}^{\alpha}\mathbf{M}_{k_3}^{\alpha}\mathbf{M}_{k_4}^{\alpha} \delta_{\mathbf{k}+\mathbf{k_2}+\mathbf{k_3}+\mathbf{k_4},0}\Bigg). 
\end{aligned}
\end{equation}

The parameters are the Ginzburg-Landau coefficients $r_0$ and $U$, the strength of the exchange $J$, Dzyaloshinsky-Moriya $D$, dipolar $\tau$ interactions, with demagnetization tensor $N$ (tr($N$) = 1), magnetic field $\mathbf{B}$ and the cubic anisotropy constant $K$.
It was already shown that the additional anisotropy term is sufficient to stabilize the LTS phase \cite{2018_Chacon_NatPhys} and to quantitatively reproduce the changes in the resonance spectra, with respect to the high-temperature skyrmion phase~\cite{2021_Back_PhysRevLett}.\\

The magnetization dynamics are described by the lossless Landau-Lifshitz equation of motion~\cite{1996_Gurevich_Book}:
\begin{equation}
\partial_t \mathbf{M} = - \gamma \mathbf{M} \times \mathbf{B}_\text{eff}
\label{eqn: LLG}
\end{equation}
Here, $\gamma = g \mu_B/\hbar$ denotes the gyromagnetic ratio and $\mathbf{B}_\text{eff} = -\delta F/ \delta \mathbf{M}$ the local effective magnetic field, given by the derivative of the total energy functional with respect to the magnetization. For further analysis, the free energy $F$ and the magnetization $\mathbf{M}$ are divided into a static and dynamic component,
\begin{equation}
\begin{aligned}
F &= F_\text{stat} + F(t)\\
\mathbf{M} &= \mathbf{M}^\text{mf} + \delta \mathbf{M}(t).
\end{aligned}
\end{equation}
Inserting this ansatz into the equation of motion \ref{eqn: LLG} and only keeping terms linear in $F(t)$ and $\delta \mathbf{M}(t)$, the resonance condition reads~\cite{2017_Waizner_PhD_2},
\begin{equation}
\omega_{\text{res}} = \text{Im} \left[\text{Eigenvalues} \big(\mathcal{W} \big)\right]
\label{eqn: Resonances}
\end{equation}
with $\mathcal{W} = \gamma \mathbf{M}^{\text{mf}} \times \chi_0^{-1}$ and $\chi_0^{-1} = \left.\frac{\delta^2 F_\text{stat}}{\delta \mathbf{M}^2}\right|_{\mathbf{M}^\text{mf}}$.\\

By performing the calculation of Eq~\ref{eqn: Resonances} in momentum space, we are able to determine the normalized eigenvectors $\textbf{\textit{v}}^\alpha_j(\mathbf{k})$ and eigenfrequencies $\omega^\alpha(\mathbf{k})$ of the given matrix $\mathcal{W}$ numerically, with internal indices $j \in [1,2,3]$ and momentum index $\mathbf{k}$.
A more detailed derivation can be found in \cite{2015_Schwarze_NatMater, 2018_Chacon_NatPhys}.\\

Based on the rather large dimensions of the CPW, we assume the driving field to be homogeneous in the performed calculations, therefore setting $\mathbf{k}$ = 0. 
The demagnetization factors of the samples under investigation read approximately $N_x = 0.17$, $N_y = 0.12$ and $N_z = 0.71$ for the platelet and $N_i = 1/3$ for a sphere, which we take as an approximation of the cuboid, where the external magnetic field is applied along the $z$-direction.\\

\begin{figure}[H]

      \centering
        \includegraphics[width = 0.75 \textwidth]{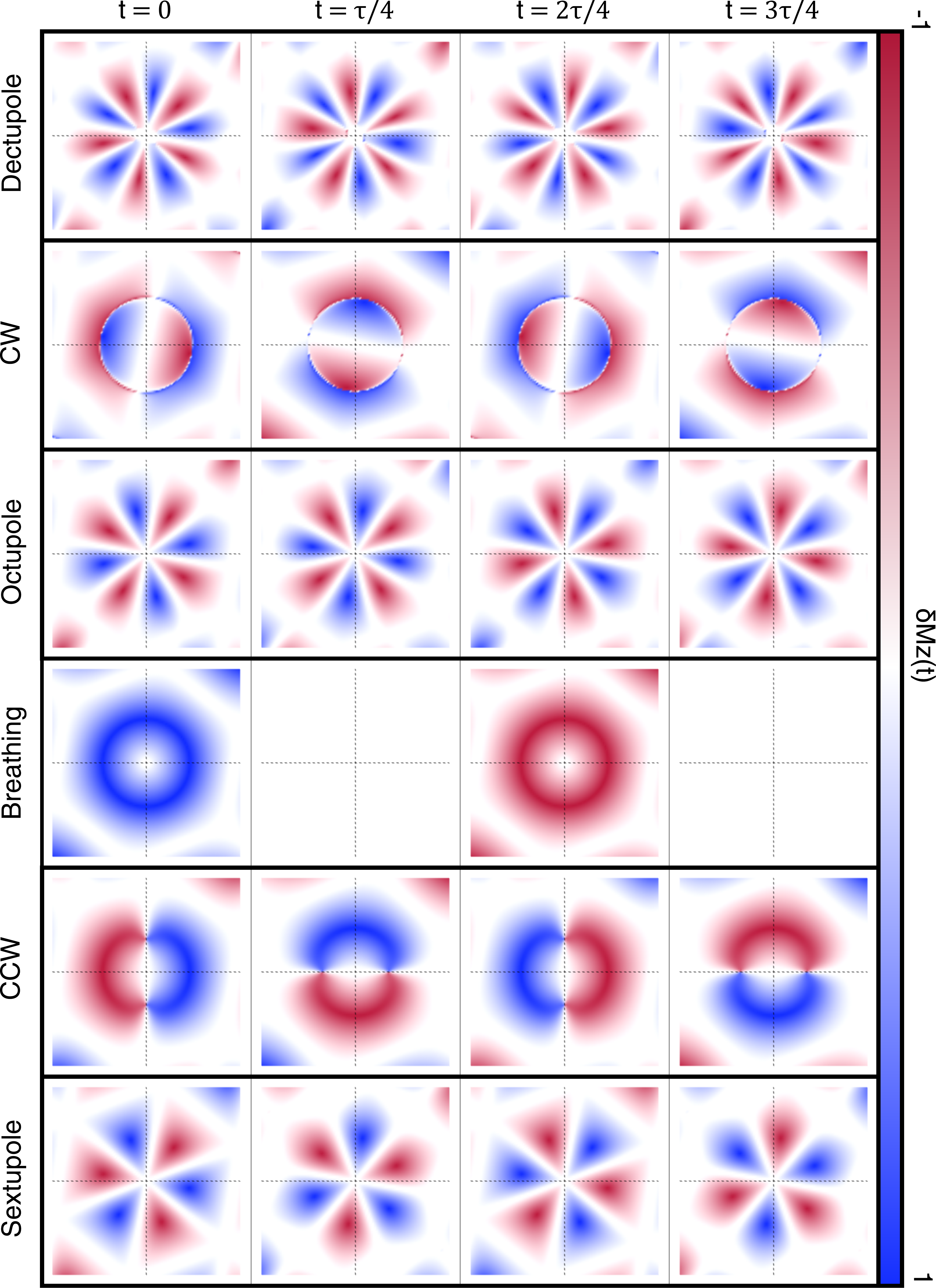}
        \caption{\textbf{Calculated real-space images of skyrmions under resonant excitation of the six lowest-frequency modes for K = 0.} The columns correspond to different characteristic moments during the oscillation; notably, they are separated in time by one-fourth of the period $\tau$. For clarity, the normalized dynamic magnetization component $\delta$$M_z$(t) is shown here. From top to bottom, the modes are identified as the dectupole, clockwise gyration mode (CW), octupole, the breathing mode (BR), counterclockwise (CCW) and the sextupole gyration.}
        \label{fig:Fig_S3}

\end{figure}

Next to the breathing, clockwise and counterclockwise mode, we focus our study on three additional modes, ranging in the same frequency regime.
For the identification of the individual modes, real space images of the normalized dynamics magnetization $\delta M_\text{z}$ are calculated for different times $t$.
The results are shown in Fig.~\ref{fig:Fig_S3}.
Based on the time evolution and the number of nodes, these modes are identified as higher-order clockwise excitations and are referred to as Sextupole (6 nodes), Octupole (8 nodes) and Dectupole(10) mode.
\\

In order to verify the repulsive character of the hybridizing modes, we calculate the inner product of the interacting resonance branches for $K = 0$ and $K \neq 0$.
For a homogeneous driving field, the internal product is given by \cite{2015_Schwarze_NatMater},
\begin{equation}
\bra{\textbf{\textit{v}}^\alpha} \ket{\textbf{\textit{v}}^\beta} :=  
\sum_{j,\mathbf{K}\in L_R} \big(\textbf{\textit{v}}^\alpha_j(\mathbf{K})\big)^\star \textbf{\textit{v}}^\beta_j(\mathbf{K})
\end{equation}
with reciprocal lattice $L_R$.
The results are presented in Table \ref{tab:Tab_2} of the main text, revealing the effect on the orthogonality of the eigenvectors of the hybridizing modes.\\\\
From our calculations, we observe that neither the shape of the sample nor a finite $\mathbf{k}$ vector changes the gap size, significantly (not shown here).
In order to further investigate the discrepancy between the experimentally and theoretically obtained gap size we added an additional anisotropy term
\begin{equation}
F_e[\mathbf{M}] = C \sum_{\mathbf{k}} \sum_{\alpha} k_{\alpha}^2 M_\mathbf{k}^\alpha M_{-\mathbf{k}}^\alpha
\label{eqn: CAnisotropy_supp}
\end{equation}
into the free energy functional. 
Depending on the sign of the prefactor $C$, the helix pitch aligns either along the $\langle111\rangle \,(C < 0)$ or the $\langle100\rangle \,(C > 0)$ direction.
In order to determine the effect of the exchange anisotropy on the hybridization gap size, we calculate the resonance spectra for various $C$ values.
For more details we refer the reader to the corresponding section in the main text.  

\clearpage

\begin{figure}[H]

        \centering
        \includegraphics[scale=0.65]{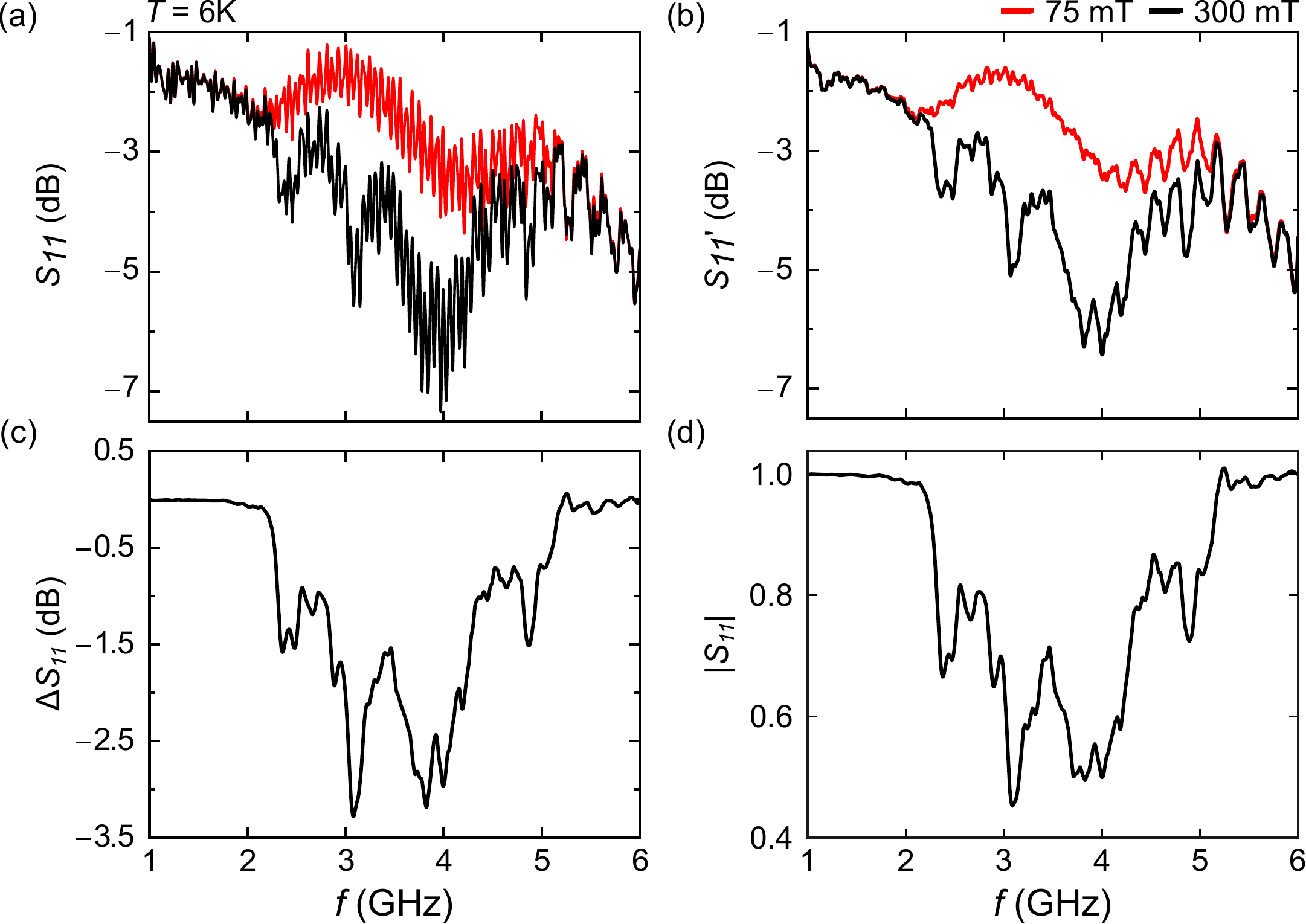}
        \caption{\textbf{Examples of microwave absorption spectra.} (a) Transmission spectra, $S_{\rm 11}$ obtained from the VNA at $\mathrm{\textit T = 6}$~K for the plate-shaped sample at $\mu_0 H$ = 75~mT and $\mu_0 H$ = 300~mT. (b) Numerically processed transmission spectra, $S_{\rm 11}'$. (c) Microwave spectra difference, $\Delta S_{\rm 11}$. (d) Microwave spectra difference after linear conversion, $|S_{\rm 11}|$.}
        \label{fig:Fig_S1}

\end{figure}

\end{document}